\documentclass[sigconf,table]{acmart}

\AtBeginDocument{%
  \providecommand\BibTeX{{%
    \normalfont B\kern-0.5em{\scshape i\kern-0.25em b}\kern-0.8em\TeX}}}

\setcopyright{none}
\renewcommand\footnotetextcopyrightpermission[1]{}
\usepackage{caption}
\usepackage{subcaption}
\usepackage{xcolor}
\usepackage{graphicx}
\usepackage{balance}
\usepackage{hyperref}
\usepackage{tcolorbox}
\usepackage[inline]{enumitem}
\usepackage{xspace}
\usepackage{cleveref}
\usepackage{multirow}
\usepackage{multicol}
\usepackage{todonotes}

\hypersetup{
	colorlinks = true,
	linkcolor = blue,
	urlcolor=blue,
	citecolor=blue
}

\begin{document}

\title{An Exploratory Study on Regression Vulnerabilities}

\author{Larissa Braz, Enrico Fregnan}
\affiliation{%
  \institution{University of Zurich}
  \country{Switzerland}
  }
\email{{larissa,fregnan}@ifi.uzh.ch}

\author{Vivek Arora}
\authornote{The author contributed to this work while being employed at Delft University of Technology, The Netherlands.}
\affiliation{%
  \institution{Independent Researcher}
  \country{India}
  }
\email{v2vivar@gmail.com}

\author{Alberto Bacchelli}
\affiliation{%
  \institution{University of Zurich}
  \country{Switzerland}
  }
\email{bacchelli@ifi.uzh.ch}


\newcommand{\etal}{\textit{et al.}\xspace}
\newcommand{\eg}{\textit{e.g.,}\xspace}
\newcommand{\ie}{\emph{i.e.,}\xspace}

\newcommand{\bg}[1]{Bug Report \texttt{#1}\xspace}
\newcommand{\cmt}[1]{Comment \texttt{#1}\xspace}

\newcommand{\totalDevsInvolved}{five\xspace}

\newcommand{\totalRegVuln}{78\xspace}
\newcommand{\totalIncorrectFixes}{72\xspace}

\newcommand{\totalSecReports}{6,667\xspace}
\newcommand{\totalAllSecRegressionReports}{187\xspace}

\newcommand{\totalReportsDefects}{78\xspace}
\newcommand{\totalReportsEnhancements}{77\xspace}
\newcommand{\totalReportsTasks}{28\xspace}
\newcommand{\totalReportsNotSet}{4\xspace}
\newcommand{\totalReportsNotSetW}{four\xspace}

\newcommand{\totalRegressionLarge}{three\xspace}
\newcommand{\totalRegressionCheck}{two\xspace}
\newcommand{\totalRegressionTiming}{seven\xspace}
\newcommand{\totalRegressionLogic}{five\xspace}
\newcommand{\totalRegressionResource}{62\xspace}

\newcommand{\totalBugLarge}{ten\xspace}
\newcommand{\totalBugCheck}{two\xspace}
\newcommand{\totalBugTiming}{six\xspace}
\newcommand{\totalBugLogic}{six\xspace}
\newcommand{\totalBugResource}{15\xspace}
\newcommand{\totalBugStructural}{34\xspace}

\newcommand{\totalFoundCrash}{15\xspace}
\newcommand{\totalFoundAssertion}{10\xspace}
\newcommand{\totalFoundASAN}{24\xspace}
\newcommand{\totalUserReported}{30\xspace}

\newcommand{\rqOne}{To what extent is software security discussed during the bug fixing process?\xspace}
\newcommand{\rqTwo}{What are the causes behind regression vulnerabilities?\xspace}
\newcommand{\rqThree}{How do developers perceive regression vulnerabilities?\xspace}

\definecolor{gray50}{gray}{.5}
\definecolor{gray40}{gray}{.6}
\definecolor{gray30}{gray}{.7}
\definecolor{gray20}{gray}{.8}
\definecolor{gray10}{gray}{.9}
\definecolor{gray05}{gray}{.95}

\definecolor{purple}{HTML}{DADAEB}
\definecolor{blue1}{HTML}{e1effc}
\definecolor{babypink}{HTML}{ffedf8}
\definecolor{palevioletred}{HTML}{db7093}

\newcommand{\interviewee}[1]{\textsc{I#1}\xspace}

\newlength\Linewidth
\def\findlength{\setlength\Linewidth\linewidth
	\addtolength\Linewidth{-4\fboxrule}
	\addtolength\Linewidth{-3\fboxsep}
}

\newenvironment{rqbox}{\par\begingroup
	\setlength{\fboxsep}{5pt}\findlength
	\setbox0=\vbox\bgroup\noindent
	\hsize=0.95\linewidth
	\begin{minipage}{0.95\linewidth}\normalsize}
	{\end{minipage}\egroup
	\textcolor{gray20}{\fboxsep1.5pt\fbox
		{\fboxsep5pt\colorbox{purple}{\normalcolor\box0}}}
	\endgroup\par\noindent
	\normalcolor\ignorespacesafterend}
\let\Examplebox\examplebox
\let\endExamplebox\endexamplebox


\newcommand{\rb}[1]{
	
	\vspace{0.3cm}
	\begin{tcolorbox}[colback=purple,
		colframe=black,
		width=\columnwidth,
		arc=3mm, auto outer arc,
		boxrule=0.5pt,
		]
		#1
	\end{tcolorbox}
}

\newcounter{Finding}
\stepcounter{Finding}

\newcommand{\roundedbox}[1]{
	\rb{
		\noindent
		\textit{\textbf{Finding \theFinding}. #1}
	}
	\stepcounter{Finding}
}

\newcommand{\vivek}[1] {\textcolor{cyan}{\textbf{(Vivek: #1)}}}
\newcommand{\lari}[1] {\textcolor{magenta}{\textbf{(Larissa: #1)}}}

\newcommand{\fix}[1] {\textcolor{red}{\textbf{(#1)}}}

\newcommand{\bluetext}[1]{\textcolor{black}{#1}}
\newcommand{\remarkEF}[1]{
	\todo[color=cyan]{\textbf{Enrico: }{#1}}
}

\newcommand{\remarkLB}[1]{
	\todo[color=yellow]{\textbf{Larissa: }{#1}}
}



\begin{abstract}
\noindent\textbf{Background:} %
Security regressions are vulnerabilities introduced in a previously unaffected software system.
They often happen as a result of source code changes (\eg a bug fix) and can have severe effects.

\noindent\textbf{Aims:}
To increase the understanding of security regressions. This is an important step in developing secure software engineering.

\noindent\textbf{Method:}
We perform an exploratory, mixed-method case study of Mozilla. 
First, we analyze \totalRegVuln regression vulnerabilities and \totalIncorrectFixes bug reports where a bug fix introduced a regression vulnerability at Mozilla.
We investigate how developers interact in these bug reports, how they perform the changes, and under what conditions they introduce regression vulnerabilities.
Second, we conduct \totalDevsInvolved semi-structured interviews with as many Mozilla developers involved in the vulnerability-inducing bug fixes. 

\noindent\textbf{Results:}
Software security is not discussed during bug fixes. Developers' main concerns are the complexity of the bug at hand and the community pressure to fix it. Moreover, developers do not to worry about regression vulnerabilities and assume tools will detect them. Indeed, dynamic analysis tools helped finding around 30\% of regression vulnerabilities at Mozilla.

\noindent\textbf{Conclusions:}
These results provide evidence that, although tool support helps identify regression vulnerabilities, it may not be enough to ensure security during bug fixes. Furthermore, our results call for further work on the security tooling support and how to integrate them during bug fixes.


\noindent\textbf{Data and materials:} \url{https://doi.org/10.5281/zenodo.6792317}

\end{abstract}



\maketitle

\newpage

\section{Introduction}
\label{intro}

Fixing bugs is an essential task for developers. 
\bluetext{Overall, developers spend 45–80\% of their time and effort looking for, understanding, and fixing bugs~\cite{collofello:1989, akbarinasaji2018predicting, zhang2012empirical}. }

However, some fixes either do not solve the problem completely or even introduce new issues~\cite{Yin:2011}, namely regression bugs~\cite{Brooks:1995}.
Regression bugs occur as an unintended consequence of program changes~\cite{Nir:2007} and happen whenever a software functionality that previously worked no longer works as planned. 
For instance, a software system may become vulnerable after a bug fix due to a vulnerability introduced by the bug fixing change itself. A software vulnerability is a security flaw, glitch, or weakness found in software code that could be exploited by an attacker~\cite{dempsey:2017} to cause harm to the stakeholders of a software system~\cite{owasp}. 

\bluetext{Mozilla~\citet{Mozilla} is a free software community that uses, develops, spreads, and supports Mozilla products, most notably the Firefox web browser. Mozilla gives significant importance to the security of its applications. For instance, one of their principles is that ``individuals’ security and privacy on the internet are fundamental and must not be treated as optional''~\cite{Mozilla:manifesto}. Yet, security vulnerabilities still affect their products and users. } 
In 2019, Mozilla rushed out a new version of Firefox to fix a critical regression vulnerability introduced with Mozilla's 12-patch security update~\cite{Networkworld-Mozilla}.
A Mozilla developer said: ``we fixed [bug] 431260, which was not really a security problem, and we introduced this bug, which probably is. Perhaps we need to be more picky about what we land on branch.'' 
Many vulnerabilities have been introduced as regressions~\cite{BleepingComputer,InfoWorld-Microsoft,washingtonpost-Apple} and put the names of big companies, such as Microsoft and Apple, under the spotlight~\cite{Yin:2011}.
To overcome this phenomenon, one should check the application's security after each source code change~\cite{Nir:2007}, including during bug fixes.

The reasons behind regression bugs can be many. For instance, \citet{Yin:2011} stated that time pressure can cause developers to not think cautiously enough while under a very tight schedule. %
Time pressure may also prevent testers from conducting thorough regression tests before releasing the fix. In addition, fixing bugs usually has a narrow focus (\eg removing the bug) compared to, \eg the development of new features. As such,
the developers fixing a bug may regard fixing the target bug as the sole objective and accomplishment to be evaluated by their manager. Therefore, they may pay significantly more attention to the bug itself than the correctness of the rest of the system.

Moreover, ideally, the developers fixing a bug should be the ones with the most knowledge about the related code. In reality, this may not always be the case. 
Nevertheless, vulnerabilities are conceptually different than traditional bugs~\cite{camilo:2015}. 
\bluetext{It is challenging to keep software systems permanently secure as changes either in the system itself or in its environment may cause new threats and software vulnerabilities~\cite{felderer:2014}. 
Thus, we need specific evidence to be able to understand and tackle regression vulnerabilities. }
To the best of our knowledge, there is still no study that investigates regression vulnerabilities, specifically how, why, and under which circumstances they are introduced. Therefore, it is necessary to investigate the circumstances that lead to the introduction of regression vulnerabilities during bug fixes, and why such fixes are marked as successful, \ie the vulnerabilities are not detected when the fixes are reviewed.

In this paper, we present an exploratory, mixed-method case study we conducted to investigate regression vulnerabilities. 
\bluetext{We focused on Mozilla as the subject of our study, because of the availability of its development data, the widespread popularity of its products, and its focus on security.}

First, we investigated the bug reports of regression vulnerabilities introduced in Mozilla through bug fixes.
We identified these cases by selecting reports marked as security issues and with their `\texttt{regressed by}' tag filled-in.\footnote{In 2018, Mozilla introduced a field (\texttt{regressed by}) into their bug tracking system Bugzilla~\cite{Bugzilla-Mozilla}. Mozilla uses the \emph{mozregression} tool~\cite{mozregression} to automatically find regression ranges. Once the change set which caused the regression has been identified, the \texttt{Regressed by} field is set to the id of the bug associated with the change-set (\ie the regression inducing bug fix). 
Every \texttt{regressed by} information is confirmed by Mozilla's developers, providing more confidence in the regression data.}
In total, we investigated \totalRegVuln regression vulnerabilities to answer why, how, and in which situations developers introduced them. We performed a qualitative analysis of both the bug fix and the regression vulnerability, including commit messages and bug reports' comments.

In the second part of our study, we conducted \totalDevsInvolved semi-structured interviews with as many Mozilla developers, who were involved in the bug fixes we investigated in the first part. Each interview lasted 30 minutes and strengthened our understanding of the reasons why regression vulnerabilities were introduced during fixes.

Our results show that \bluetext{a total of \totalIncorrectFixes bug fixes introduced the \totalRegVuln regression vulnerabilities we investigated. Security was not a concern in the comments of these bug fixes}: Developers only discussed the potential security impact of the bug fix in 5 out of the \totalIncorrectFixes reports. 
Concerning regression vulnerabilities, our findings highlight that most vulnerabilities were caused by structural or resource defects: 34 bug fixes impacted the structure or the organization of the code \bluetext{(\eg moved a functionality from one module to another)}, while 16 concerned the \bluetext{misuse of memory resources (\eg buffer copy without checking input's size - CWE-120~\cite{CWE:online})}. The primary source of detection of regression vulnerabilities is Mozilla's users (30 out of the 79 analyzed regression vulnerabilities). However, dynamic analysis tools, \bluetext{such as address sanitizers~\cite{Mozilla:sanitizers}}, also played a fundamental role in detecting regression vulnerabilities. The interviewees confirmed the importance of the tools, yet recognized their limitations and need for improvement.


\section{Background and Related Work}
\label{related}

We \bluetext{describe the Mozilla’s bug fixing process and} summarize relevant literature on software security and the bug fixing process.

\subsection{\bluetext{Mozilla's bug fixing process}}
\noindent\textbf{\bluetext{Bug reports:}}
As its main bug tracking system, Mozilla uses Bugzilla, a web-based general-purpose bug tracking system~\cite{Bugzilla-Mozilla}.
Bugs are organized by type to make it easier to (1) perform triage, (2) assign a bug to a developer, and (3) understand a release's quality. A bug report can be labeled as a `\texttt{defect}' (regression, crash, security vulnerability, and any other reported issue), `\texttt{enhancement}' (new feature, UI improvement, and any other request for user-facing enhancements to the product, not engineering changes), or `\texttt{task}' (refactoring, removal, replacement, and any other engineering task). 
All types need triage decisions from engineers or product managers. \bluetext{Mozilla states that distinguishing between defects and tasks is important to understand code quality and reduce defects introduced while working on new features or fixing existing defects~\cite{Mozilla:bugTypes}. }

\smallskip
\noindent\textbf{\bluetext{Regressions:}}
Mozilla defines a regression as a `\texttt{defect}' introduced by a change~\cite{mozilla-regressions}. For instance, for regression bugs reported in Mozilla-Central (the main development tree for Firefox), the policy is to tag the bug as a regression, identify the commits causing the regression, then mark the bugs associated with those commits as causing the regression.
Mozilla uses the \emph{mozregression} tool~\cite{mozregression} to find regression ranges. 
The tool uses a binary search algorithm for quickly determining the changeset range corresponding to when a problem was introduced. Until the changeset causing the regression has been found through \emph{mozregression} or another bisection tool, the bug is also tagged as `\texttt{regression-window-wanted}'. Once the changeset which caused the regression is identified, the previous keyword is removed and the `\texttt{Regressed by}' field is set to the id of the bug associated with the changeset. Setting the `\texttt{Regressed by}' field updates the `\texttt{Regresses}' field in the regression-inducing bug. In addition, a `\texttt{needinfo}' tag is set for the author of the regressing patch, asking them to fix/revert the change. Mozilla's developers confirm every `\texttt{Regressed by}' value. 

\smallskip
\noindent\textbf{\bluetext{Software vulnerabilities:}}
Mozilla considers a bug as a \emph{security bug} when it has been reported as \texttt{security-sensitive} in Bugzilla and received a security rating. According to Mozilla guidelines~\cite{mozilla_private}, all security bugs should be reported as private, making the process for patching them slightly different than the usual process for fixing a bug. A security-sensitive bug in Bugzilla means that all its information, except its ID number, are hidden. This includes the title, comments, reporter, assignee, and CC'd people. 

A security-sensitive bug usually remains private until a fix is shipped in a new release and after a time to ensure that a maximum number of users updated their version of Firefox. Bugs are usually made public after six months and a couple of releases.
From the moment a security bug has been privately reported to when a fix is shipped and the bug is set public, all information about that bug is handled carefully to avoid an unmitigated vulnerability becoming known and exploited before Mozilla releases a fix (0-day).

Moreover, Mozilla asks contributors to be careful not to disclose sensitive information about the bug in public places, such as Bugzilla. They should not add public bugs in the \texttt{duplicate}, \texttt{depends on}, \texttt{blocks}, \texttt{regression}, \texttt{regressed by}, or \texttt{see also} sections if these bugs could give hints about the nature of the security issue. In our study, we analyze only the public security vulnerabilities tagged as regressions in Mozilla's Bugzilla. 

\newpage
\subsection{\bluetext{Related work}}

\noindent\textbf{Incorrect bug fixes: }%
Despite its omnipresence, code change is perhaps the least understood and most complex aspect of the development process~\cite{purushothaman:2004}. However, code changes, such as bug fixes, are not bulletproof as they are written by human~\cite{Yin:2011}. Some fixes either do not solve the problem entirely or even introduce new ones. 

Previous studies have investigated incorrect bug fixes~\cite{Sliwerski:2005,purushothaman:2004,gu:2010,baker:1994,Yin:2011}. For instance, \citet{Sliwerski:2005} proposed a way to automatically locate fix-inducing changes and studied the incorrect fix ratios in Eclipse and Mozilla. 
They identified that developers make more bug fixing mistakes on Fridays.

\citet{Yin:2011} presented a characteristic study on incorrect bug-fixes from four large operating systems, investigating not only the mistake patterns during bug fixes but also the possible human reasons in the development process when these incorrect fixes were performed. They found that at least $14.8\%\sim24.4\%$ of fixes for post-release bugs are incorrect and impact end users. Developers and reviewers involved in incorrect fixes usually do not have enough knowledge about the code under analysis. For example, $27\%$ of the incorrect fixes are made by developers who have never touched the files associated with the fix. However, traditional bugs are conceptually different from software vulnerabilities~\cite{camilo:2015}. Therefore, evidence about the former may not translate to the latter. For this reason, further studies are needed to investigate regression vulnerabilities introduced by incorrect bug fixes.

\smallskip
\noindent\textbf{Security regression detection: } %
Keeping software systems permanently secure is challenging: changes either in the system itself or in its environment may introduce new threats and software vulnerabilities~\cite{felderer:2014}. Regression testing is performed to ensure that changes made to existing software do not cause unintended effects on unchanged parts of the software and that the changed parts behave as intended~\cite{chen:2002, leung:1989}. Therefore, a combination of regression and security testing, called security regression testing, which ensures that changes made to a system do not harm its security, is of high significance~\cite{Felderer2015}. \citet{mehta:2007} underlined the importance of regression testing while verifying the vulnerabilities of a system.

Furthermore, \citet{Felderer2015} identified a retest-all approach as the most commonly applied regression testing technique for regression vulnerability detection. However, they also stated how retest-all has significant limitations in terms of fault detection and localization. In our study, we further investigate whether and how developers ensure security during bug fixes in practice. For example, we study whether developers use regression detection tools when fixing the bugs or reviewing the fix changes.

\section{Methodology}
\label{lab:methodology}

Our goal is to gain a deeper knowledge on regression vulnerabilities, in particular how and why they are introduced. We run a case study based on a mixed-method approach to collect different types of qualitative and quantitative evidence.

\subsection{Research Questions}

We structured our study around three main research questions. 
First, we focus on understanding the circumstances around the introduction of regression vulnerabilities. Thus, we investigate whether security is a developer's concern during bug fixes:


\begin{center}
	\begin{rqbox}
		\begin{description}	\item[]\textbf{RQ$_1$.} \emph{\rqOne}
		\end{description} 
	\end{rqbox}
\end{center}

After assessing the role of software security in the bug-fixing process, we aim to understand the causes that led to the introduction of regression vulnerabilities. 
Thus, we ask:

\smallskip

\begin{center}
	\begin{rqbox}
		\begin{description}	\item[]\textbf{RQ$_2$.} \emph{\rqTwo}       
		\end{description} 
	\end{rqbox}
\end{center} 

Finally, we investigate developers' perceptions of regression vulnerabilities through semi-structured interviews.

\smallskip

\begin{center}
	\begin{rqbox} 
		\begin{description}
			\item[]\textbf{RQ$_3$.} \emph{\rqThree}           	                
		\end{description} 
	\end{rqbox}
\end{center}

\begin{figure}[t]
	\centering
	\includegraphics[width=1\columnwidth]{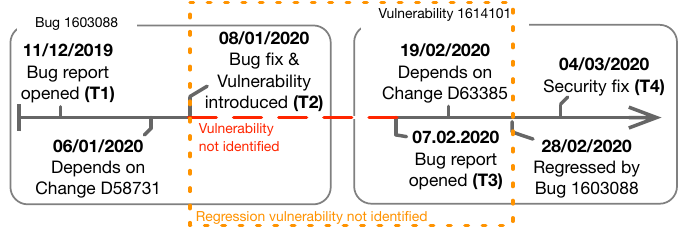}
	\caption{Timeline of a regression vulnerability in Mozilla.} 
	\label{fig:timeline}
\end{figure}

\subsection{Vulnerability-Inducing Bug Fixes Analysis}
\label{sec:met:bugreportanalysis}

\bluetext{\Cref{fig:timeline} presents the timeline of a regression vulnerability in Mozilla. We see that a bug report 1603088 was opened on 11/12/2019 (T1) and later fixed on 06/01/2020 (T2). However, the bug fix introduced a vulnerability in the code. On 07/02/2020, a new bug report 1614101 was opened to report a vulnerability (T3). It was identified as a regression of the first bug fix only on 28/02/2022 (T3). Finally the vulnerability was fixed on 04/03/2020 (T4).}

To study regression vulnerabilities that happened in the history of Mozilla, we manually analyzed (1) the bug reports labeled as \texttt{defects} that introduced regression vulnerabilities \bluetext{(\eg Bug 1603088 in \Cref{fig:timeline})} and (2) the bug reports of the security issues tagged as regressions \bluetext{(\eg Vulnerability 1614101 in \Cref{fig:timeline})}.

To retrieve the regression software vulnerabilities, we queried Bugzilla with the settings: \texttt{Keywords:} \texttt{sec-high}, \texttt{sec-critical, sec-moderate, sec-incident}, \texttt{sec-low}, and \texttt{Regressed by: (is not empty)}. Subsequently, we retrieved information on bugs inducing vulnerabilities by accessing the bug reports referenced in the `\texttt{Regressed by}' field of the vulnerabilities' reports. We filtered out the reports labeled as \texttt{enhancements} or \texttt{tasks}; instead, we analyzed only the bug reports labeled as \texttt{defects}, as well as the reports of the vulnerabilities they regressed.

In our manual analysis, we first analyzed the report of the bug that introduced the vulnerability (the vulnerability-inducing bug, Bug 1614101 in \Cref{fig:timeline}) and also the report of the regression vulnerability (Vulnerability 1614101 in \Cref{fig:timeline}) by going through their descriptions and titles provided by the bug reporters, and the discussion comments left in the reports by the developers. 
We performed an open card sort~\cite{spencer2009card} to extract emerging themes from the comments of the bug reports of both the vulnerability-inducing bug (1603088?) and the regression vulnerability (1614101). The first author created self-contained units, then sorted them into themes. The units were interactively sorted several times to ensure the themes' integrity. The last author reviewed the final themes. Through discussions around the themes, we evaluated controversial answers, reduced potential bias caused by wrong interpretations of comments from the bug reports, and strengthened the confidence in the card sorting output (available in our replication package~\cite{replication-package}).

Moreover, we applied the General Defect Classification scheme (GDC)~\cite{Emam:1998,Mantyla:2009} to classify (1) the bugs, (2) the bugs' fixing code changes (T2 in \Cref{fig:timeline}), and (3) the regression vulnerabilities. The GDC scheme classifies code review comments, static analysis tool warnings, technical comments or suggestions, and code analysis. Due to the broader scope of the bug reports, GDC can help deriving meaningful results from the manual analysis. 

To complete our investigation, we classified the collected regression vulnerabilities using the Common Weakness Enumeration (CWE) taxonomy~\cite{CWE:online}. CWE is a comprehensive list of common programming errors that can lead to a vulnerability. In addition, CWE's documentation provides details about the security issues, such as description, relationship to other CWEs, consequences of the security weakness, examples of code snippets, common attack patterns, and notes related to its maintenance and possible research gaps. 
The current version of CWE list includes $922$ security weaknesses. Due to its cause-based categorization, it can help identify and prevent common programming  errors that might lead developers to introduce vulnerabilities in the application's code.

\subsection{Semi-structured Interviews}
\label{sec:met:interviews}

To investigate the developers' perspective on regression vulnerabilities, we performed a set of one-to-one semi-structured interviews~\cite{lindlof:2002} with software developers at Mozilla.
Semi-structured interviews rely on interview guidelines containing general groupings of topics and questions rather than a strict pre-determined set of questions~\cite{lindlof:2002,Bacchelli:2013}. Therefore, the interviewer can adjust the interview flow based on the answers. In semi-structured interviews, participants can share their thoughts and perceptions while providing researchers with the opportunity to explore new ideas as they emerge during the interview~\cite{hiller2004,hove2005}. Given these characteristics, semi-structured interviews are often used in exploratory studies to ``find out what is happening [and] to seek new insights''~\cite{weiss:1995}. 

Semi-structured interviews help us collect developers' descriptions, experiences, observations, and assessments of how Mozilla developers  deal with regression vulnerabilities. The semi-structured interviews complement the previously conducted analysis of bug reports as these may not completely reflect the bug fixing process: \eg developers may not describe the reasoning behind all decisions taken during the bug fix. 

The interview guideline was collectively developed and revised several times by the authors of this paper. 
The main path of the final interview guideline has eleven questions. 
In the following, we describe the main topics within our guideline. 

\smallskip
\noindent\textbf{(1) Introduction and background.} We start each interview spending a few minutes explaining who we are and our research. We review the main points of the consent form interviewees signed prior to the interviews and answer any  questions. 
Subsequently, we ask questions about the participants' backgrounds (\eg their programming experience and current role). This discussion serves as an icebreaker and provides us with some context.

\smallskip
\noindent\textbf{(2) Bug fixing.} In this part of the interview, we ask general questions about the bug fixing process. First, we ask participants to explain the bug fixing process at Mozilla. Although Mozilla has publicly available documentation on the processes followed by its developers, teams and individual developers can adjust these guidelines to suit their needs or the products they are developing. 
Moreover, we ask participants their definition of regression during fixes and whether specific tests or guidelines exist for regressions.

\smallskip
\noindent\textbf{(3) Regression vulnerabilities.} In this step, we introduce the topic of regressions vulnerabilities and ask specific questions about it. We aim to understand how participants deal with this type of regression and whether they are any different from other types of regressions (\eg performance regressions). In addition, we also inquire about the methods and practices the interviewees follow to avoid introducing regression vulnerabilities while fixing a bug. 

\smallskip
\noindent\textbf{(3) Advice for novices.} We ask participants about the main challenges they face when dealing with regression vulnerabilities. We also discuss what has proven most helpful during these activities. 
Finally, we ask what advice they would give novice developers regarding security during bug fixes and regression vulnerabilities.

\smallskip
\noindent\textbf{(4) Wrap-up.} We conclude the interviews with wrap-up questions to collect final feedback from the participants. 

\smallskip
We conducted the interviews through a video-conferencing application. After obtaining the participants' consent, we recorded the interviews. The audio of each interview was then transcribed and divided into smaller coherent units for subsequent analysis. 
Moreover, we performed card sorting~\cite{spencer2009card} to extract emerging themes from the interviews. 
We followed a similar process as in the analysis of the bug reports (see \Cref{sec:met:bugreportanalysis}): The first and second authors created self-contained units, then sorted them into themes. To ensure the themes' integrity, the authors interactively sorted the units several times until the final themes were reached. Through discussions among the authors involved in this process, we evaluated controversial answers, reduced potential bias caused by wrong interpretations of a participant's answer, and strengthened the confidence in the card sorting output. 
The complete interview guideline, the transcripts of the interviews, and the card sorting outputs are available in our replication package~\cite{replication-package}.\smallskip

\noindent\textbf{Recruiting Participants.}
To recruit participants for the interviews, we contacted software developers from Mozilla who reported a vulnerability-inducing bug (T1) or a software vulnerability (T3), as well as the developers assigned as bug maintainers of these reports. We also contacted developers at Mozilla in the authors' personal network.
For each interview participant, we donated 30 USD to the Mozilla Foundation or a charity chosen by the interviewee as a token of appreciation for their time and effort.
\section{Results}
\label{results}

In this section, we report the results of our study.

\subsection{Regression Vulnerabilities and Bug Reports}

In total, Bugzilla returned \totalSecReports public bug reports of security issues when we applied the keywords filter (\Cref{sec:met:bugreportanalysis}). By filtering these vulnerability with the key \texttt{Regressed by: (is not empty)}, we retrieved \totalAllSecRegressionReports regression vulnerabilities reports (T3 in \Cref{fig:timeline}).

Then, we proceeded to retrieve the bug reports (T1 in \Cref{fig:timeline}) of the fixes that introduced the vulnerabilities using the bug \texttt{id} listed in the \texttt{Regressed by} field of the regression vulnerabilities  reports. In total, \totalReportsDefects regression vulnerabilities were introduced by bug reports labeled as \texttt{defects}, \totalReportsEnhancements by reports labeled as enhancements, and \totalReportsTasks by reports labeled as tasks. This label was blank or not set in \totalReportsNotSetW reports. In our study, we only consider the vulnerabilities regressed by bug reports labeled as \texttt{defects}. 
Moreover, seven bug fixes introduced more than one regression vulnerability at a time. In these cases, we considered the vulnerability-inducing bug reports and their fix changes (T2 in \Cref{fig:timeline}) only once in our analysis. 
After applying these exclusion criteria, we could use data from \totalRegVuln regression vulnerability reports and \totalIncorrectFixes vulnerability-inducing bug reports for the analysis. 

\begin{figure}[b]
	\centering
	\includegraphics[width=0.95\columnwidth]{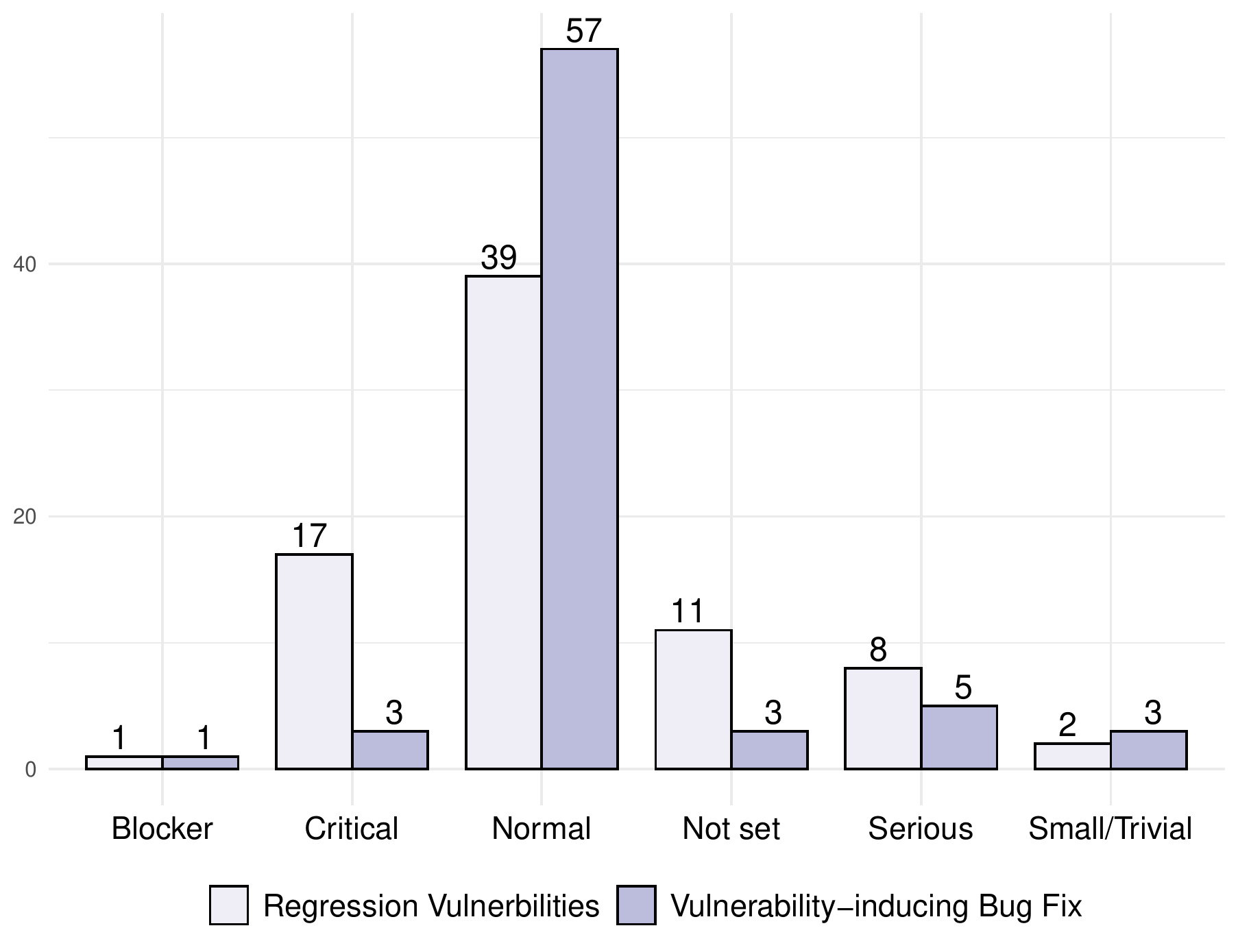}
	\caption{Severity of the vulnerability-inducing bug fixes and the regression vulnerabilities.} 
	\label{fig:severity}
\end{figure}

\begin{figure}[b]
	\centering
	\includegraphics[width=0.7\columnwidth]{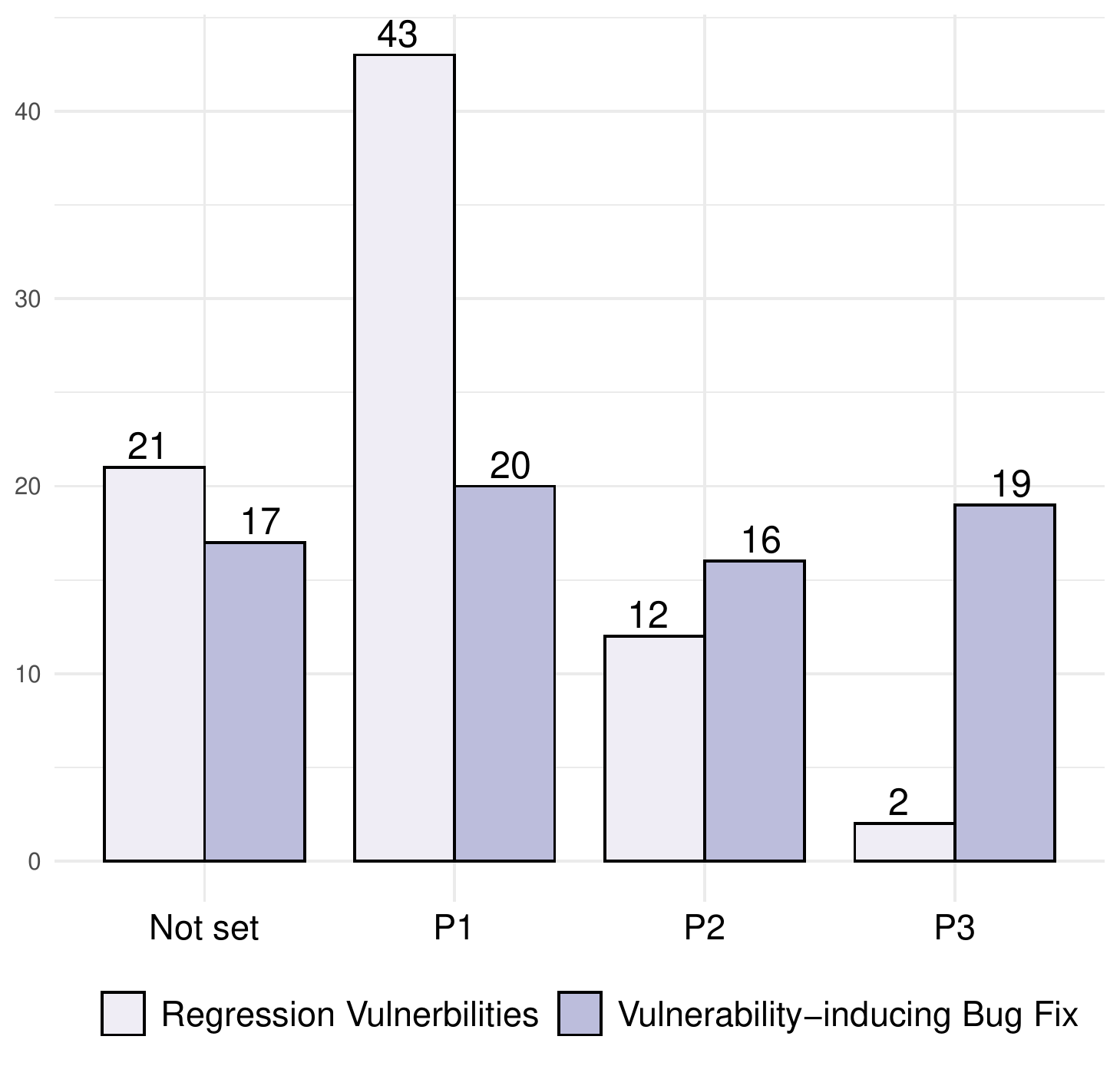}
	\caption{Priority of the vulnerability-inducing bug fixes and the regression vulnerabilities.} 
	\label{fig:priority}
\end{figure}

Among the \totalRegVuln regression vulnerabilities considered in our study, 68 occurred in Mozilla Core, 5 in the Toolkit, 3 in Firefox, 1 in the NSS project, and 1 in the GeckoView project. In total, 39 regression vulnerabilities were defined as normal severity (\ie they blocked non-critical functionality and a workaround existed), 25 regression vulnerabilities were defined as critical or serious (\ie a major functionality/product was severely impaired and a workaround did not exist), two vulnerabilities were defined as small or trivial (\ie cosmetic issues, low or no impact to users), one vulnerability was defined as a blocker (\ie blocked the development/testing and might have impacted more than 25\% of the users, caused data loss, potential chemspill, and no workaround was available), and 11 regression vulnerabilities did not have their severity defined. 
\Cref{fig:severity} shows the severity of the analyzed regression vulnerabilities.

Moreover, most regression vulnerabilities (43) were prioritized as P1, which means they should be fixed in the current release cycle. In total, 12 vulnerabilities were P2, \ie to fix in the next or following release cycle, two were set as P3 (backlog), and 21 did not have their priority set. No regression vulnerability was set as ``will not fix, but accept a patch.'' \Cref{fig:severity} presents the priorities of the regression vulnerabilities analyzed in our study. 
The bug reports of the regression vulnerabilities remained open for a median of 11 days. Most regression vulnerabilities' reports (53) were closed in less than 20 days, six of them were closed in one day or less. We also detected a case of a regression vulnerability closed only after 137 days. 

\bluetext{%
Considering the bug fixes that led to these vulnerabilities, we found that 57 had normal severity, eight were critical or serious bugs, three were trivial or minor bugs, one was a catastrophic bug, and three did not have a set severity level. The priority of the bugs was almost evenly distributed: 20 were P1; 16 were P2; 19 were P3; and 17 of them did not have a set priority. Figures \ref{fig:severity} and \ref{fig:priority} show the severity and priority of the analyzed vulnerability-inducing bug fixes, respectively.}

\bluetext{%
The analyzed bug reports were open between February 2007 and December 2020 and took a median of 37.5 days to be closed. In total, 6 bug reports were closed in one day or less, while 23 bug reports remained open for more than 90 days. Nine bug reports remained open for more than 1,100 days. }

\subsection{\bluetext{RQ$_1$. Security in the Bug Fixing Process}}

\bluetext{Our first research question investigates whether security is taken into account during the bug fixing process, looking at bug fixes that led to the introduction of regression vulnerabilities.}

The bug reports had a median of 17 comments from 10 different authors (bots excluded). A bug report that spent 36 days open had the highest number of comments (214) made by 18 different authors. Three bug reports had only three comments, and two of them were closed in one day or less. Finally, seven bug reports contained only bot comments. 

We manually analyzed the comments in the bug reports. \emph{Only in five bug reports} the commenters expressed a concern about the security impacts of the bug fixes. For instance, Comment 27 of \bg{471015} stated: ``Fixing that would require more or less a complete rewrite of the fieldset code. Frame-splitting code tends to be very fragile and a great source of security bugs. Do we really want to land that rewrite on stable branches?'' In addition, \cmt{27} in \bg{1450965} requested security experts to review the bug fix: ``if we want to land on possible solution in comment 11, we will need someone from security team to take a look if the proposed solutions are ok.''
In total, four bug reports had comments indicating that the bug might have been security-related. For instance, in \bg{1513201}, \cmt{16} said: ``[Triage Comment] Fixes a security-sensitive crash, approved for 65.0b5.'' 

Moreover, commenters pressured the maintainers for fixes in four bug reports. For instance, \bg{471015} remained \texttt{3,990} days opened and 57 developers left 73 comments. \cmt{46} said: ``If this is the way, Mozilla wants to handle this problem, maybe this bug should be declared as `wont fix.' Otherwise, a fixing date would be very kind. I know that there might be more important things to do, but \dots this bug is going to be 5 years old!?!'' \cmt{49} said: ``I don't intend to be nasty but this bug is open for way too long (I've open the same myself in 2003 with \bg{191308}) and nobody seems to care to correct it --> this is *MORE* than 10 years!'' Security was not addressed in any comment of this bug. 

Nine reports were reopened after the regression vulnerability was identified, indicating that their bug fixes were updated to remove the security flaw in the code. The remaining regression vulnerabilities were fixed through their own changesets. In addition, ten bug reports are regressions caused by previous bug fixes. In other three bug reports, comment authors demonstrated concern about the possibility of the fix being a regression. For instance, the reporter of \bg{1459905} said: ``[Regression range]: This seems to be an old regression, I was not able to reproduce it on FF 4.0, I will follow up with a regression range as soon as possible.''

A total of nine bug reports had comments stating the bugs were non-trivial to understand or fix. For instance, \cmt{24} of \bg{471015} mentioned: ``Based on bug [report] 463350 Comment 17, it sounds like our (currently-disabled) frameset-breaking code is pretty broken, and a safe fix-up would require some non-trivial rewriting, which could be scary to land on branches (and could potentially open us up).'' 
In addition, \cmt{2} of \bg{1557765} said: ``This bug has a pretty high technical level and I do not have the skill to understand it or confirm it. I will set its component as (Core) Performance and hope that a [developer] can have a more appropriate opinion on it.''

Moreover, \cmt{34} of \bg{371787} reports that a trick would be needed to ensure the correctness of the code: ``Some trickiness would be needed to ensure getEndPositionOfChar and getExtentOfChar don't include any letter-spacing or word-spacing that comes after it. I haven't handled that \dots I'm inclined to leave that to a follow-up.'' 
This bug stayed open for 4,650 days and had 51 comments from 36 different authors. The comments in the report included it not being a trivial bug to fix (``It's not trivial to get working.'' --- \cmt{7}) and a discussion about hiring a software developer to fix the bug (``How likely are you to submit a patch to implement or pay someone to do so on your behalf, [Author X]?'' and ``Who can do it? - I would like to spend some money to solve this issue. Maybe we can share the costs??''--- Comments 14 and 16, respectively). After more than five years, the bug report was closed and reopened again as it was identified as the cause of a regression vulnerability: ``Yes, it was disabled in bug 1599173 because of a serious regression, and then re-enabled when fixed.''

Although they introduced regression vulnerabilities in the code, some authors of bug fixes were initially acknowledged for their work. For instance, \cmt{33} of \bg{1231213} complimented the work done in its patch: ``Truly fantastic work on this entire patch set. The future looks very bright for ServiceWorkers in Gecko!'' However, this bug fix introduced a regression vulnerability of normal severity with P1 priority that took 50 days to be fixed. In addition, \cmt{4} of \bg{1052579} said: ``It's a good tactic. :-) I used it in bug 991981 five years ago. It can be a bit of a sprawling tactic, but that's how the cookie crumbles. :-| [at] least you aren't doing it for a security bug.'' In this case, the bug fix introduced a critical regression vulnerability that was fixed within seven days after being reported. 

\vspace{-0.1cm}
\roundedbox{Security was rarely a concern among the comments in the issue reports of bugs whose fixing introduced vulnerability regressions.}\smallskip

\subsection{RQ$_2$. Causes of regression vulnerabilities}
\label{sec:results:rq2}

\begin{figure}[b]
	\centering
	\includegraphics[width=0.9\columnwidth]{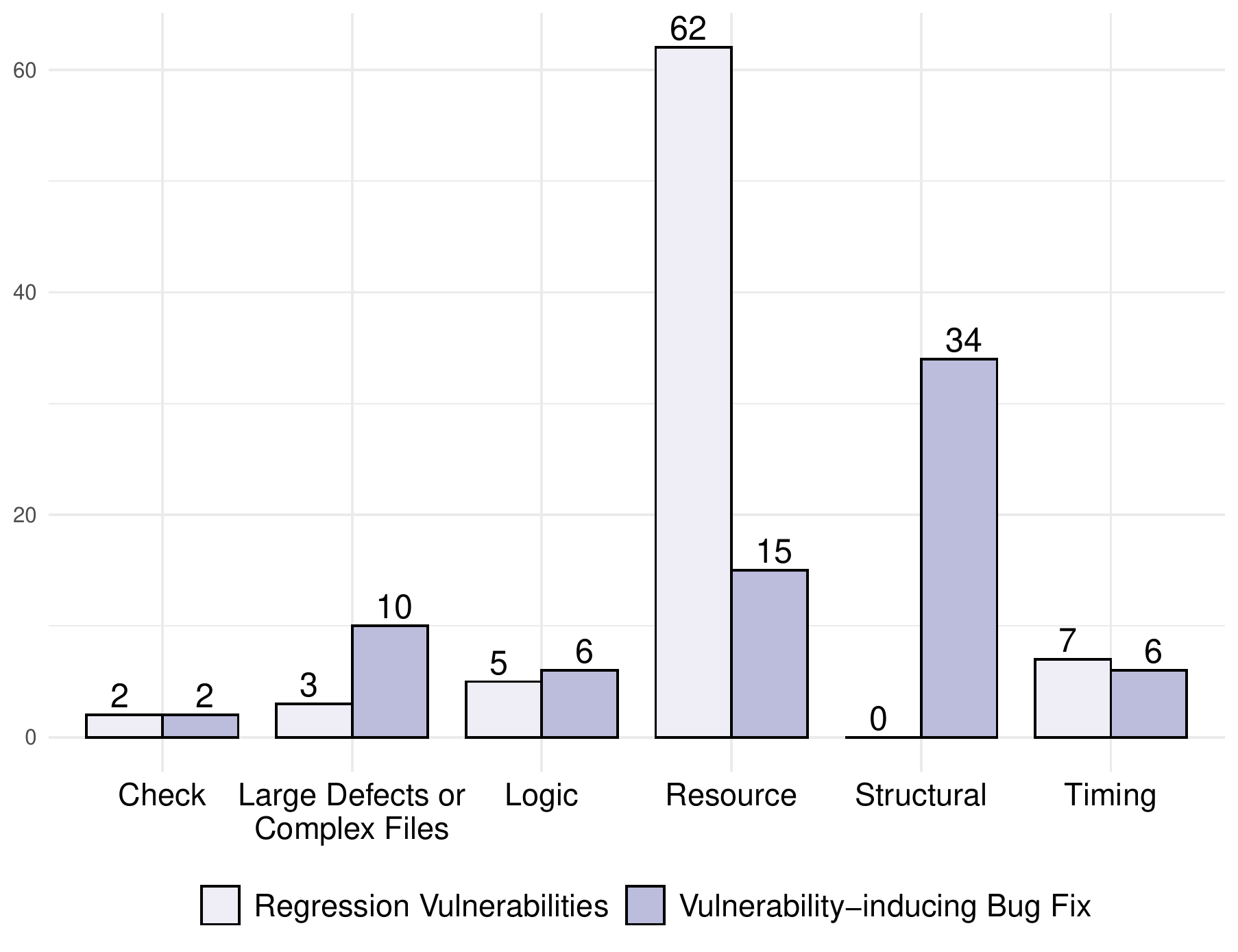}
	\caption{Causes of the vulnerability-inducing bug fixes and regression vulnerabilities.} 
	\label{fig:distribution}
\end{figure}

We applied GDC~\cite{Emam:1998,Mantyla:2009} to manually classify the changes made in the \totalIncorrectFixes vulnerability-inducing bug fixes considered in our study. We classified the bug fixes into six categories: structural, resource, large defects or complex files, timing, logic, and check defects. \Cref{fig:distribution} shows the results of our analysis.

In total, we classified \totalBugCheck bug fixes as check defects. These defects result from missing or incorrect data validation, such as parameters, simple or complex variables, and return values. The logic category of defects results from logical errors in computation or the control flow. The difference between check and logic defects is that former validates incorrect or correct values (null checks or true and false conditions), while the latter looks for correct comparisons (A is greater or less than B). We classified \totalBugLogic bugs as caused by logic defects. 

In total, \totalBugTiming bug were caused by timing defects, which happen when multiple threads operate on shared resources. These defects include race conditions and protecting resources with synchronization primitives.
In addition, the resource category includes defects that happened while handling resources (\eg memory buffer, sessions, and connections). Developers create custom resources according to the requirements; therefore, defects introduced while handling such data structures are also part of the resource category. In total, we classified \totalBugResource bug fixes as resource defects. For \bg{1530178}, its bug was due to both logic and resource defects. 
 
Moreover, we identified \totalBugStructural bug fixes as structural defects. This category of defects includes the organization of code. For instance, a developer may introduce a bug by moving a functionality from one module to another, removing duplicate or complex code, long subroutine, or dead code. 
In addition, we also considered the complexity of the bug fix. Large defects that contains bug reports with multi-part fixes or where the fixes are distributed across multiple files. This is because we note that \totalBugLarge bug fixes that introduced security regressions are also either large or complex in nature.

We further investigated the causes of the regression vulnerabilities to identify the developers' mistakes during the bug fixes that introduced them. We identified 116 flaws in the vulnerability-inducing bug fixes, which we classified into 26 Common Weaknesses Enumeration (CWEs). In total, 34 regression vulnerabilities have \textit{use after free} (CWE-416) flaws in the code. This CWE has a low base score (4.7), \ie a low sum of the impact and exploitability scores, and happens when the code references memory after it has been freed causing a program to crash, use unexpected values, or execute code. The second most frequent security flaw in the bug fixes was \textit{improper check or handling of exceptional conditions} (CWE-703, 16). This vulnerability occurs when the code does not properly anticipate or handle exceptional conditions that rarely occur during the normal operation of the application. Finally, the third most common vulnerabilities were \textit{buffer copy without checking size of input} (CWE-120, 7) and \textit{improper control of a resource through its lifetime} (CWE-664, 7). The former happens when the code performs operations on a memory buffer reading from or writing to a memory location that is outside of the intended boundary of the buffer. The latter occurs when the software does not properly follow instructions on how to create, use, and destroy resources leading to unexpected behaviors and exploitable states.
\Cref{tab:cwedistribution} shows the CWEs causing the regression vulnerabilities we analyzed.

\begin{table}[t]
\centering 
\caption{Categories of the regression vulnerabilities according to the Common Weakness Enumeration (CWE).} 
\label{tab:cwedistribution}
\begin{tabular}{llr}
\hline
\textbf{CWE} & \textbf{Name}                         & \textbf{Count} \\ 
\hline
CWE-416      & Use after free    & 34                              \\ 
CWE-703      & Improper check or handling & 16                              \\ 
                       & of exceptional conditions & \\
CWE-120       & Buffer copy without checking size & 7\\
						& of input  &                             \\
CWE-664      & Improper control of a resource            & 7                               \\ 
						&  through its lifetime & \\
CWE-119       & Improper restrictions of operation & 6                               \\
                       & within the bound of  a memory buffer &  \\
CWE-662      & Improper synchronization                                      & 5                               \\ 
CWE-911       & Improper update of reference count                            & 5                               \\ 
CWE-665      & Improper initialization                                       & 5                               \\ 
CWE-131       & Incorrect calculation of buffer size                           4 &                                \\
CWE-20         & Improper input validation                                          & 4                               \\ 
CWE-366      & Race condition within a thread                                              & 4                               \\ 
CWE-200      & Exposure of sensitive information                                   & 4                               \\ 
					& to an unauthorized actor & \\
CWE-79         & Improper neutralization of input during & 3 \\
 						& web page generation            &                                \\ 
CWE-730      & Denial of service                                                           & 2                               \\ 
CWE-177       & Improper handling of URL encoding                                         & 2                               \\ 
CWE-451       & User Interface (UI) misrepresentation   & 2                               \\
						&  of critical information      &                    \\ 
CWE-465       & Pointer issues                                         & 1                               \\ 
CWE-470       & Use of externally-controlled input & 1 \\
&  to select class or code                                            &                          \\ 
CWE-185        & Incorrect regular expression                                  & 1                               \\ 
CWE-130       & Improper handling of length                   & 1                               \\ 
                       & parameter inconsistency& \\
CWE-190       & Integer overflow or wraparound                                              & 1                               \\ 
CWE-227       & API Abuse                                                     & 1                               \\ 
CWE-522       & Insufficiently protected credentials   & 1                               \\ 
CWE-415       & Double free                                                   & 1                               \\ 
CWE-241       & Improper handling of unexpected  & 1                               \\ 
                        & data type & \\
CWE-228       & Improper handling of syntactically  & 1                               \\ 
                       & invalid structure & \\
\hline
\end{tabular}
\end{table}

\vspace{-0.2cm}
\roundedbox{Structural and resource defects are prevalent causes of bugs. Whereas most of the regression vulnerabilities induced by the fixes of these bugs occur due to defects related to resources.}\smallskip

\bluetext{Knowledge on the causes of defects and security vulnerabilities might support practitioners in preventing these defects from happening. For instance, projects might adopt technologies less prone to defects related to memory resources.}

Following, we investigated how the regression vulnerabilities were detected and reported. In total, 30 of the regression vulnerabilities were found and reported by Mozilla's users. For instance, the reporter of \bg{1670358} said: ``We came across this when we were working on something similar to a ChaosMode feature where we wanted to make array reallocation happen more often.'' This vulnerability was caused by a use after free flaw (CWE-416) introduced seven years before by the fix of the \bg{877762}. It was reported as CVE-2020-26960 with a high base score (9.3).

\bluetext{The remaining vulnerabilities were identified by dynamic analysis tools (24) (\eg address sanitizers~\cite{Mozilla:sanitizers}), crashes due to user usage (15), and assertion failures (10). }
\bluetext{Concerning the detection of vulnerabilities from the users,} Mozilla provides the crash reports through their `crash-stat' website~\cite{crash-stat}. In addition, Mozilla also allows users to custom build the artifacts with switches such as \texttt{-enable-}\texttt{debug} and \texttt{enable-crash-on-assert} to enable additional logging and crash dump generation. For instance, Mozilla's \texttt{mozilla-central} code repository can be built with the \texttt{-enable-} \texttt{debug} switch to enable the debug logging. This way, maintainers can obtain more knowledge about a defect, facilitating its fixing. 
 
\vspace{-0.2cm}
\roundedbox{Dynamic analysis tools played an essential role in finding the analyzed regression vulnerabilities.}

\subsection{RQ$_3$. Developers' perception on regression vulnerabilities}

We interviewed \totalDevsInvolved Mozilla software developers. 
Two of them are security experts, while the others have management or development roles. 
Moreover, participants have between 10 and 24 years of experience contributing to Mozilla. On average, they commented on 8,146 bug reports, which indicates they were involved in their bug fixes. Participants were assigned to an average of 919 bug reports. In these cases, they were directly involved in the bug fixes. Finally, they reviewed on average 1,684 patches in Mozilla's repositories. 
\Cref{tab:intervieweesdem} summarizes the background of the interviewees.

The interviews' card sorting analysis revealed 100 cards of 9 main categories and 59 sub-categories. We considered each card only once per interview.
Overall, all developers discussed security concerns and the practices adopted during bug fixes at Mozilla (15 cards). 
\bluetext{Additional findings emerged that were not envisioned as part of the focus of our study. However, they are still important to report. For instance, }security bugs are more complex to be fixed than other types of bugs: \eg they are harder to identify using tests as writing security-specific tests can expose a potential weakness to malicious attackers. 
\interviewee{2} mentioned: ``[security] is a lot more complex than general bugs. We cannot just add a test that says we will check for this vulnerability because that will paint a target on the change.''

\begin{table}[b]
	\centering 
	\caption{Demographics of the interviews participants.} 
	\label{tab:intervieweesdem}
	\begin{tabular}{llrrrr}
		\hline
		\multicolumn{1}{c}{\textbf{ID}} & \textbf{Role} & \multicolumn{1}{c}{\textbf{C. Years}} & \textbf{R.C.} & \textbf{A.T.} & \textbf{P.R.} \\ 
		\hline
		\emph{I1} & Security engineer & 24 & 22,910 & 2,504 & 6,703 \\
		\emph{I2} & Maintainer & 10 & 3,797 & 715 & 442 \\
		\emph{I3} & Engineering manager & 18 & 2,682 & 347 & 750 \\
		\emph{I4} & Developer \&& 11 & 4,607 & 607 & 287 \\
 &   Engineering manager&  &  &  &  \\
		\emph{I5} &Security engineer \& & 12 & 6,738 & 422 & 242 \\
		& Tech lead & &  &  &  \\
		\hline
		\multicolumn{6}{l}{{\small C. years = Years contributing to Mozilla; R.C. = Bug reports the}} \\
		\multicolumn{6}{l}{{\small participant commented on; A.T. = Bug reports assigned to}} \\
		\multicolumn{6}{l}{{\small the participant; P.R. = Reviewed patches.}} \\
	\end{tabular}
\vspace{-0.3cm}
\end{table}

Moreover, developers also mentioned the use of security bugs as a way to mentor junior developers. For instance, \interviewee{1} stated: ``We use security bugs as a good mentoring tool for junior developers. The senior developers may not have the time to do the fix themselves, but it gives them an option \dots to teach that part of the code to maybe some new hires we have or more junior people to solve it and then they learn the kind of the security impacts and assumptions that happen in that part of the code.'' 
However, this training may lead to potentially harmful assumptions when ensuring security during bug fixes. Since junior developers receive security training, the security team does not worry too much about fixing patches but focuses on patches introducing new features, as reported by \interviewee{1}.

\roundedbox{Interviewees stated that security bugs are more complex to be fixed than other types of bugs. Security bugs are used as a mentoring tool to train junior developers.}

Developers also mentioned the role of tests in ensuring security during bug fixes (10 cards). \interviewee{2} reported that adding tests is required during the bug fix process when test coverage is too low. In addition, when asked what their team does to identify possible regression vulnerabilities during bug fixes, \interviewee{4} replied: ``mostly review and the automated testing, so the developers review each other's patches trying to identify if there are possibilities for regressions and then after the patch is actually accepted and everything. We test them automatically and manually to ensure that there are no regressions.''

However, three interviewees (\interviewee{2}, \interviewee{3}, and \interviewee{5}) highlighted that regressions may happen despite tests as they may not cover all platforms or all possible code interactions.  
For instance, \interviewee{5} said: ``There is always this type of regression that only reproduces in very certain environments under very certain conditions (...) that are really not possible for us to easily find in in an automated way.'' 
To mitigate this issue, Mozilla has a Quality Assurance team that periodically manually runs a list of regression tests (as mentioned by \interviewee{4}). Moreover, \interviewee{1} reported that each part of Mozilla's code has a designated set of people with security experience that reviews and approve code changes. 

\vspace{-0.2cm}
\roundedbox{Tests are required during bug fixes at Mozilla. Yet, tests may not detect regressions as they can not cover complex interaction scenarios.}\smallskip

Developers also discussed the tool support to detect vulnerabilities (26 cards). For instance, \interviewee{2} said: ``We have extensive tooling to avoid [regression vulnerabilities]. So, when we run tests, we have things like address sanitizers, static analysis that will flag potential problems before the patch goes into review.'' 
Static analysis tools that developers at Mozilla use to identify vulnerabilities are also employed to avoid regression vulnerabilities during bug fixes (\interviewee{1}, \interviewee{3}, \interviewee{4}, and \interviewee{5}). Despite the benefits of these tools, security vulnerabilities can still go undetected: \eg \interviewee{4} reported: ``We have the static analysis tools but we will not find them preventing so many security issues.''

Improving analysis tools would further reduce the number of regression vulnerabilities (2 cards). \interviewee{4} reported: ``I think there is a lot of room for improvements in [the static analysis] area to identify common [security] mistakes.'' 
However, developers could not provide any concrete suggestions on how to improve these tools. As \interviewee{3} said: ``I'm not aware of much that we could add at this point that we have not already tried.''

In total, we identified 11 cards about fuzzing. Four interviewees mentioned fuzzing as a way of identifying vulnerabilities in Mozilla's code. For instance, \interviewee{4} reported: ``[Fuzzing] is another thing we do that is super helpful (...) to prevent a lot of security bugs.''
Moreover, \interviewee{1} mentioned a specific fuzzing testing team that reviews bug fixes so that they do not introduce regression vulnerabilities: ``so depending on the bug, I have a fuzz testing team that is looking for security bugs specifically.'' 
The advantages of fuzzing go beyond detecting vulnerabilities, as fuzzing provides developers with information to better understand the root cause of code issues. 

\interviewee{5} highlighted the efficiency of fuzzing in finding regression vulnerabilities: ``Other way we find regressions really early is fuzzing. Fuzzing is a very effective method for finding regressions, simply because once you have fuzzing in place, you can automate it very easily.'' However, fuzzing may be a costly activity, as mentioned by \interviewee{1}: ``It makes the builds run slower, so we cannot ship it that way, but it does catch those kinds of bugs sooner.'' Fuzzers do not run for every bug fix at Mozilla (as reported by \interviewee{4} and \interviewee{5}) but only periodically to check for defects in the code.

Developers also mentioned the importance of using Address Sanitizers (ASANs) to detect use after free and out-of-bound vulnerabilities in C/C++ code. For instance, \interviewee{5} mentioned: ``If we would remove ASANs from our testing, we would have significantly less detection rate for memory corruption vulnerabilities. Like when address sanitizer was introduced into our testing workflow six to eight years ago \dots we found a ton of longstanding use after free bugs that would otherwise be really hard to detect because they do not necessarily manifest these crashes or have any other side effects if they are not being exploited right now. So you could say that deploying certain types of sanitizers is primarily for detecting more security vulnerabilities.'' 
In our previous analysis, we found that \textit{use after free} caused 43\% of the investigated regression vulnerabilities (see \Cref{sec:results:rq2} and \Cref{tab:cwedistribution}). In addition, 30\% of the regression vulnerabilities were detected through ASANs (see \Cref{sec:results:rq2}). 

\vspace{-0.2cm}
\roundedbox{Tools play an important role in the detection of software vulnerabilities at Mozilla. However, tools need to be improved to better support developers in avoiding to introduce regression vulnerabilities during bug fixes.}\smallskip

Two interviewees stated that Mozilla has no guidelines to avoid introducing regression vulnerabilities during bug fixes. Despite not having specific security guidelines, vulnerabilities are still considered more severe than functional regressions. This way, they receive more visibility and are investigated more quickly (\interviewee{4}). 
Moreover, how to deal with a regression vulnerability in Mozilla's code depends on its impact. \interviewee{2} reported: ``If the regression has a very large impact, we are going to back out the change that caused the regression instead of fixing the bug. We do that first, then we fix the original bug again, and this time we do it correctly. In the case of a normal bug, that does not happen, obviously.'' 

\roundedbox{Mozilla does not have a guideline to avoid introducing regression vulnerabilities during bug fixes. Yet, they are considered more severe than other type of regressions.}

\section{Limitations}
In this section, we report the limitations of this study and how we mitigated them. 

Our analysis was limited only to regression vulnerabilities that happened at Mozilla. Mozilla is a popular open-source project with a vast number of active contributors that follow state-of-the-art software development practices. However, despite these characteristics, Mozilla might not represent other development contexts. For instance, the culture of a company may influence what happens during the bug fixing process (\eg through the use of company-specific guidelines). For this reason, we cannot exclude that a similar study conducted in a different company might lead to different results. 

In our investigation, we considered only \textit{public} regression vulnerabilities returned by our query of Mozilla's Bugzilla as well as the bug fixes linked as the cause of these regressions. The bug tracking system might include \textit{private} regression vulnerabilities, which are not publicly available for external investigations. Nonetheless, we cannot exclude that including these regression vulnerabilities in our analysis may lead to different results. 

Moreover, the \emph{mozregression} tool (that Mozilla uses to identify the bug fix that caused a regression) may give inaccurate results. To mitigate this issue, Mozilla developers need to manually review the information provided by the tool and approve it. This strengthens our confidence in the fact that the evaluated bug fixes indeed introduced regression vulnerabilities. 

To mitigate \emph{mono-method bias}~\cite{Cook:1979}, we collected data from both bug reports and interviews with software developers and used card-sorting to identify common themes. 
In the interviews, we followed guidelines to avoid biasing the participants' answers. Nonetheless, we cannot exclude that the interview settings might have influenced participants to provide more desirable answers~\cite{hildum:1956}. To mitigate this bias, we challenged and triangulated our findings with the results of the analysis of the bug reports.
Furthermore, interview participants might have given socially acceptable answers to appear in a positive light. To mitigate this social desirability bias~\cite{furnham:1986}, we informed participants that the responses would been anonymized and evaluated in a statistical form.  

To obtain a diverse sample of participants, we invited Mozilla developers from several teams and roles with different education levels and development experience. Despite the diversity of the participants' backgrounds, we cannot claim that our sample is representative of all Mozilla developers. This might influence the generalizability of our findings. 

Participants could freely decide whether to participate in the interviews or not (self-selection). They were informed about the interview’s topic and its estimated duration. Moreover, to encourage developers' participation we donated to a charity institution on their behalf. This could have biased the participants' selection as only participants who could spare enough time or were interested in the incentive might have participated. 
\section{Discussion}

In this section, we discuss how our findings on Mozilla’s security practices during bug fixes as well as developers' perception of regression vulnerabilities can inform research and practice.

\vspace{0.10cm}
\noindent
\textbf{Collaboration is key for security.} %
Mozilla has extensive tooling in place to support developers in detecting security issues in the code. In addition, developers are required to write tests during the bug fixing process to identify potential defects introduced by the fix changes. Yet, despite the benefits of these practices, these alone may not be enough to detect all possible security issues introduced by the changes. In fact, even security experts can have difficulties to detect vulnerabilities as they need to identify the portions of the code where to focus their testing and inspection efforts~\cite{shin:2010}. Interview participants mentioned Mozilla's multi-platform support, and code interactions and complexity as possible reasons behind regression vulnerabilities. Indeed, \citet{shin:2010} studied two open-source projects, including Mozilla, and found evidence that complexity can make code difficult to understand and test for security. 

To help developers \bluetext{avoid regression vulnerabilities}, \textit{collaboration} between software developers and security experts is fundamental (as reported by the interview participants). Developers should work together with a mentor and have stronger collaborations with the security and testing teams. 
%
%
\citet{Braz:2021} reported how security training with a mentor could strengthen developers' security knowledge. 
Yet, to learn security, developers must be able to switch from their traditional conditioning to the attacker’s way of thinking~\cite{Bratus:2007}. Junior developers might be paired with a security expert during bug fixes to learn and improve the security of fix changes. For instance, at Mozilla, security bugs are used as a mentoring tool to teach security to junior developers. 
Further study should be conducted to investigate whether and how security knowledge transfer can happen in the bug fixing process. Moreover, approaches should be devised to better support this knowledge exchange. 



\vspace{0.10cm}
\noindent
\textbf{Never drop your guard.} %
\bluetext{The results for RQ$_1$ provide evidence that security is rarely a concern among the comments in bug reports at Mozilla.} 
Even though junior developers at Mozilla receive security training, regression vulnerabilities still happen in practice. Interview participants reported that the security team assumes this training reduces the risk of vulnerabilities being introduced during bug fixes. Therefore, they are usually not too worried about investigating patches for typical vulnerabilities. 
However, recent research~\cite{Braz:2021, braz:2022} provided initial evidence that security is not routinely in the mind of developers when reviewing code: \eg reviewing changes in the bug-fixing process. This might be the reason why regression vulnerabilities still occur despite developers' security training. 
In addition, even though developers report caring more about security than reliability issues~\cite{Christakis:2016}, only a limited number of them adopt secure coding practices~\cite{woon:2007}. 
For these reasons, companies should reinforce their security policies and security teams should remain alert for potential vulnerabilities in the code, not overestimating the effects of security training for non-experts.

The presence of tools may give developers a false sense of security, decreasing their attention to potential vulnerabilities introduced in the code. For instance, developers at Mozilla reported assuming that tools will detect most vulnerabilities introduced during bug fixes. However, in our study, only ~30\% of the analyzed regression vulnerabilities were identified by analysis tools (RQ$_2$). Thus, analysis tools should not be considered as the solution for software vulnerabilities. Tools should complement, rather than replacing, developers' training and attention to security. 
Companies and team leads should foster a security-oriented development culture, emphasizing the key role that developers play in guaranteeing the security of the code and investing in training and guidelines to support them in writing secure applications.

\vspace{0.10cm}
\noindent
\textbf{Choosing the right tech stack.} Interview participants reported that to ensure the application's security, developers need to know and understand the security risks in the tech stack (\eg programming language) being used. 
A tech stack is the combination of technologies a company uses to build and run an application or project~\cite{heap-techstack}. It typically consists of programming languages, frameworks, a database, front-end tools, back-end tools, and applications connected via APIs. 
Choosing the appropriate tech stack is of paramount importance to increase the security of a project. For instance, Mozilla created the Oxidation project to integrate Rust code in and around Firefox~\cite{oxidation-mozilla}. They state that roughly 70\% of critical security vulnerabilities are caused by memory safety bugs and crashes, which are considerably less likely to happen using Rust. 
\bluetext{Indeed, most of the  regression vulnerabilities (RQ$_2$) occurred due to defects related to memory resources, such as buffer copy without checking size of input (CWE-120).} 
We suggest that, when designing applications, architects should work closely with security experts to select secure technologies for the project. At the same time, efforts need to be spend on informing developers about the security aspects of the project's tech stack.  For instance, to improve software security in a project, software engineers might receive training on the security risks of the technology they use.

\section{Conclusions}

In this paper, we performed an exploratory mixed-method study to investigate the characteristics of regression vulnerabilities: how, why, and when these regressions happen as well as how developers perceive them.
To this aim, we inspected \totalIncorrectFixes Mozilla's reports of bugs whose fixing introduced security issues and \totalRegVuln reports of regression vulnerabilities. We also conducted semi-structured interviews with \totalDevsInvolved Mozilla developers involved in the vulnerability-inducing bug fixes to better understand their perception and practices regarding regression vulnerabilities. 

Our findings indicate that security is not a concern in the comments of vulnerability-inducing bug fixes: Developers discussed security aspects only in 6\% of the analyzed bug reports. The interviews revealed that avoiding regression vulnerabilities is not the developers' goal during the bug fixing process. They rely on tools to detect such issues before they cause problems to the application. 

Our results show that most vulnerability-inducing bug fixes were caused by structural or resource defects. Moreover, most of them were detected by Mozilla's users. Yet, dynamic analysis tools also played a fundamental role in detecting regression vulnerabilities. 

Finally, our investigation raises questions on the effectiveness of currently used methods to ensure security in the software development process, especially during bug fixing. To achieve and maintain secure applications, companies and project owners should adopt a more security-oriented culture: \eg choosing a tech stack that better supports security and strengthening collaboration between developers and security experts. 

\begin{acks}
The authors would like to thank the anonymous reviewers for their thoughtful and important comments, which helped improving our paper. The authors gratefully acknowledge the support of the Swiss National Science Foundation through the SNSF Projects No. 200021\_197227 and PZ00P2\_186090.
\end{acks}

\newpage
\bibliographystyle{ACM-Reference-Format} 
\bibliography{references}


\begin{thebibliography}{52}


\ifx \showCODEN    \undefined \def \showCODEN     #1{\unskip}     \fi
\ifx \showDOI      \undefined \def \showDOI       #1{#1}\fi
\ifx \showISBNx    \undefined \def \showISBNx     #1{\unskip}     \fi
\ifx \showISBNxiii \undefined \def \showISBNxiii  #1{\unskip}     \fi
\ifx \showISSN     \undefined \def \showISSN      #1{\unskip}     \fi
\ifx \showLCCN     \undefined \def \showLCCN      #1{\unskip}     \fi
\ifx \shownote     \undefined \def \shownote      #1{#1}          \fi
\ifx \showarticletitle \undefined \def \showarticletitle #1{#1}   \fi
\ifx \showURL      \undefined \def \showURL       {\relax}        \fi
\providecommand\bibfield[2]{#2}
\providecommand\bibinfo[2]{#2}
\providecommand\natexlab[1]{#1}
\providecommand\showeprint[2][]{arXiv:#2}

\bibitem[Again(2022)]%
        {washingtonpost-Apple}
\bibfield{author}{\bibinfo{person}{Apple Updates~Leopard Again}.}
  \bibinfo{year}{Last accessed March 2022}\natexlab{}.
\newblock \bibinfo{title}{Washington Post}.
\newblock
\newblock
\newblock
\shownote{\url{http://voices.washingtonpost.com/fasterforward/2008/
  02/apple_updates_leopardagain.html}}.


\bibitem[Akbarinasaji et~al\mbox{.}(2018)]%
        {akbarinasaji2018predicting}
\bibfield{author}{\bibinfo{person}{Shirin Akbarinasaji}, \bibinfo{person}{Bora
  Caglayan}, {and} \bibinfo{person}{Ayse Bener}.}
  \bibinfo{year}{2018}\natexlab{}.
\newblock \showarticletitle{Predicting bug-fixing time: A replication study
  using an open source software project}.
\newblock \bibinfo{journal}{\emph{journal of Systems and Software}}
  \bibinfo{volume}{136} (\bibinfo{year}{2018}), \bibinfo{pages}{173--186}.
\newblock


\bibitem[Bacchelli and Bird(2013)]%
        {Bacchelli:2013}
\bibfield{author}{\bibinfo{person}{A. Bacchelli} {and} \bibinfo{person}{C.
  Bird}.} \bibinfo{year}{2013}\natexlab{}.
\newblock \showarticletitle{Expectations, outcomes, and challenges of modern
  code review}. In \bibinfo{booktitle}{\emph{Proceedings of the International
  Conference on Software Engineering}}. \bibinfo{pages}{712--721}.
\newblock


\bibitem[{Baker} and {Eick}(1994)]%
        {baker:1994}
\bibfield{author}{\bibinfo{person}{M. {Baker}} {and} \bibinfo{person}{S.
  {Eick}}.} \bibinfo{year}{1994}\natexlab{}.
\newblock \showarticletitle{Visualizing software systems}. In
  \bibinfo{booktitle}{\emph{Proceedings of the International Conference on
  Software Engineering}}. \bibinfo{pages}{59--67}.
\newblock


\bibitem[Bratus(2007)]%
        {Bratus:2007}
\bibfield{author}{\bibinfo{person}{S. Bratus}.}
  \bibinfo{year}{2007}\natexlab{}.
\newblock \showarticletitle{What Hackers Learn that the Rest of Us Don't: Notes
  on Hacker Curriculum}.
\newblock \bibinfo{journal}{\emph{IEEE Security Privacy}} \bibinfo{volume}{5},
  \bibinfo{number}{4} (\bibinfo{year}{2007}), \bibinfo{pages}{72--75}.
\newblock


\bibitem[Braz et~al\mbox{.}(2022a)]%
        {braz:2022}
\bibfield{author}{\bibinfo{person}{Larissa Braz}, \bibinfo{person}{Christian
  Aeberhard}, \bibinfo{person}{G{\"u}l {\c{C}}alikli}, {and}
  \bibinfo{person}{Alberto Bacchelli}.} \bibinfo{year}{2022}\natexlab{a}.
\newblock \showarticletitle{Less is More: Supporting Developers in
  Vulnerability Detection during Code Review}. In
  \bibinfo{booktitle}{\emph{2022 IEEE/ACM 44th International Conference on
  Software Engineering (ICSE)}}. IEEE, \bibinfo{pages}{1317--1329}.
\newblock


\bibitem[Braz et~al\mbox{.}(2022b)]%
        {replication-package}
\bibfield{author}{\bibinfo{person}{L. Braz}, \bibinfo{person}{E. Fregnan},
  \bibinfo{person}{V. Arora}, {and} \bibinfo{person}{A. Bacchelli}.}
  \bibinfo{year}{2022}\natexlab{b}.
\newblock \bibinfo{title}{Data and Material}.
\newblock \bibinfo{howpublished}{\url{https://doi.org/10.5281/zenodo.6792317}}.
\newblock


\bibitem[Braz et~al\mbox{.}(2021)]%
        {Braz:2021}
\bibfield{author}{\bibinfo{person}{Larissa Braz}, \bibinfo{person}{Enrico
  Fregnan}, \bibinfo{person}{G{\"u}l {\c{C}}alikli}, {and}
  \bibinfo{person}{Alberto Bacchelli}.} \bibinfo{year}{2021}\natexlab{}.
\newblock \showarticletitle{Why Don’t Developers Detect Improper Input
  Validation?{'; DROP TABLE Papers;--}}. In \bibinfo{booktitle}{\emph{2021
  IEEE/ACM 43rd International Conference on Software Engineering (ICSE)}}.
  IEEE, \bibinfo{pages}{499--511}.
\newblock


\bibitem[Brooks(1995)]%
        {Brooks:1995}
\bibfield{author}{\bibinfo{person}{F. Brooks}.}
  \bibinfo{year}{1995}\natexlab{}.
\newblock \bibinfo{booktitle}{\emph{The mythical man-month: essays on software
  engineering}}.
\newblock \bibinfo{publisher}{Pearson Education}.
\newblock


\bibitem[Camilo et~al\mbox{.}(2015)]%
        {camilo:2015}
\bibfield{author}{\bibinfo{person}{F. Camilo}, \bibinfo{person}{A. Meneely},
  {and} \bibinfo{person}{M. Nagappan}.} \bibinfo{year}{2015}\natexlab{}.
\newblock \showarticletitle{Do bugs foreshadow vulnerabilities? A study of the
  chromium project}. In \bibinfo{booktitle}{\emph{Proceedings of the Working
  Conference on Mining Software Repositories}}. \bibinfo{pages}{269--279}.
\newblock


\bibitem[Chen et~al\mbox{.}(2002)]%
        {chen:2002}
\bibfield{author}{\bibinfo{person}{Y. Chen}, \bibinfo{person}{R. Probert},
  {and} \bibinfo{person}{D.l Sims}.} \bibinfo{year}{2002}\natexlab{}.
\newblock \showarticletitle{Specification-based regression test selection with
  risk analysis}. In \bibinfo{booktitle}{\emph{Proceedings of the conference of
  the Centre for Advanced Studies on Collaborative research}}.
\newblock


\bibitem[Christakis and Bird(2016)]%
        {Christakis:2016}
\bibfield{author}{\bibinfo{person}{M. Christakis} {and} \bibinfo{person}{C.
  Bird}.} \bibinfo{year}{2016}\natexlab{}.
\newblock \showarticletitle{What developers want and need from program
  analysis: an empirical study}. In \bibinfo{booktitle}{\emph{Proceedings of
  the international conference on automated software engineering}}.
  \bibinfo{pages}{332--343}.
\newblock


\bibitem[Collofello and Woodfield(1989)]%
        {collofello:1989}
\bibfield{author}{\bibinfo{person}{J. Collofello} {and} \bibinfo{person}{S.
  Woodfield}.} \bibinfo{year}{1989}\natexlab{}.
\newblock \showarticletitle{Evaluating the effectiveness of
  reliability-assurance techniques}.
\newblock \bibinfo{journal}{\emph{Journal of systems and software}}
  \bibinfo{volume}{9}, \bibinfo{number}{3} (\bibinfo{year}{1989}),
  \bibinfo{pages}{191--195}.
\newblock


\bibitem[Computer(2022)]%
        {BleepingComputer}
\bibfield{author}{\bibinfo{person}{Bleeping Computer}.} \bibinfo{year}{Last
  accessed March 2022}\natexlab{}.
\newblock \bibinfo{title}{Windows 10 gets temp patch for critical flaw fixed in
  buggy update}.
\newblock
\newblock
\newblock
\shownote{\url{https://www.bleepingcomputer.com/news/security/windows-10-gets-temp-patch-for-critical-flaw-fixed-in-buggy-update/}}.


\bibitem[{Cook} and {Campbell}(1979)]%
        {Cook:1979}
\bibfield{author}{\bibinfo{person}{T. {Cook}} {and} \bibinfo{person}{D.
  {Campbell}}.} \bibinfo{year}{1979}\natexlab{}.
\newblock \bibinfo{booktitle}{\emph{Quasi-Experimentation: Design and Analysis
  Issues for Field Settings}}.
\newblock \bibinfo{publisher}{Houghton Mifflin Company}.
\newblock


\bibitem[Dempsey et~al\mbox{.}(2017)]%
        {dempsey:2017}
\bibfield{author}{\bibinfo{person}{K. Dempsey}, \bibinfo{person}{P. Eavy},
  {and} \bibinfo{person}{G. Moore}.} \bibinfo{year}{2017}\natexlab{}.
\newblock \bibinfo{booktitle}{\emph{Automation Support for Security Control
  Assessments}}.
\newblock \bibinfo{type}{{T}echnical {R}eport}. \bibinfo{institution}{Technical
  Report NISTIR 8011, National Institute of Standards and Technology}.
\newblock


\bibitem[El~Emam and Wieczorek(1998)]%
        {Emam:1998}
\bibfield{author}{\bibinfo{person}{K. El~Emam} {and} \bibinfo{person}{I.
  Wieczorek}.} \bibinfo{year}{1998}\natexlab{}.
\newblock \showarticletitle{The repeatability of code defect classifications}.
  In \bibinfo{booktitle}{\emph{Proceedings Ninth International Symposium on
  Software Reliability Engineering (Cat. No.98TB100257)}}.
  \bibinfo{pages}{322--333}.
\newblock


\bibitem[Felderer and Fourneret(2015)]%
        {Felderer2015}
\bibfield{author}{\bibinfo{person}{M. Felderer} {and} \bibinfo{person}{E.
  Fourneret}.} \bibinfo{year}{2015}\natexlab{}.
\newblock \showarticletitle{{A systematic classification of security regression
  testing approaches}}.
\newblock \bibinfo{journal}{\emph{International Journal on Software Tools for
  Technology Transfer}} \bibinfo{volume}{17}, \bibinfo{number}{3}
  (\bibinfo{year}{2015}), \bibinfo{pages}{305--319}.
\newblock


\bibitem[Felderer et~al\mbox{.}(2014)]%
        {felderer:2014}
\bibfield{author}{\bibinfo{person}{M. Felderer}, \bibinfo{person}{B. Katt},
  \bibinfo{person}{P. Kalb}, \bibinfo{person}{J. J{\"u}rjens},
  \bibinfo{person}{M. Ochoa}, \bibinfo{person}{F. Paci}, \bibinfo{person}{T.
  Tun}, \bibinfo{person}{K. Yskout}, \bibinfo{person}{R. Scandariato},
  \bibinfo{person}{F. Piessens}, {et~al\mbox{.}}}
  \bibinfo{year}{2014}\natexlab{}.
\newblock \showarticletitle{Evolution of security engineering artifacts: a
  state of the art survey}.
\newblock \bibinfo{journal}{\emph{International Journal of Secure Software
  Engineering}} \bibinfo{volume}{5}, \bibinfo{number}{4}
  (\bibinfo{year}{2014}), \bibinfo{pages}{48--98}.
\newblock


\bibitem[Foundation(2017)]%
        {mozregression}
\bibfield{author}{\bibinfo{person}{Mozilla Foundation}.}
  \bibinfo{year}{2017}\natexlab{}.
\newblock \bibinfo{title}{mozregression tool}.
\newblock
  \bibinfo{howpublished}{\url{https://mozilla.github.io/mozregression/}}.
\newblock


\bibitem[Foundation(2022h)]%
        {oxidation-mozilla}
\bibfield{author}{\bibinfo{person}{Mozilla Foundation}.} \bibinfo{year}{Last
  accessed April 2022}\natexlab{h}.
\newblock \bibinfo{title}{Oxidation Project}.
\newblock \bibinfo{howpublished}{\url{https://wiki.allizom.org/Oxidation}}.
\newblock


\bibitem[Foundation(2022d)]%
        {mozilla-regressions}
\bibfield{author}{\bibinfo{person}{Mozilla Foundation}.} \bibinfo{year}{Last
  accessed in March 2022}\natexlab{d}.
\newblock \bibinfo{title}{How to Mark Regressions}.
\newblock
  \bibinfo{howpublished}{\url{https://firefox-source-docs.mozilla.org/bug-mgmt/processes/regressions.html}}.
\newblock


\bibitem[Foundation(2022a)]%
        {Mozilla:manifesto}
\bibfield{author}{\bibinfo{person}{Mozilla Foundation}.} \bibinfo{year}{Last
  accessed June 2022}\natexlab{a}.
\newblock \bibinfo{title}{Address Manifesto}.
\newblock
  \bibinfo{howpublished}{\url{https://www.mozilla.org/en-US/about/manifesto/}}.
\newblock


\bibitem[Foundation(2022b)]%
        {Mozilla:sanitizers}
\bibfield{author}{\bibinfo{person}{Mozilla Foundation}.} \bibinfo{year}{Last
  accessed June 2022}\natexlab{b}.
\newblock \bibinfo{title}{Address Sanitizers}.
\newblock
  \bibinfo{howpublished}{\url{https://firefox-source-docs.mozilla.org/tools/sanitizer/index.html}}.
\newblock


\bibitem[Foundation(2022c)]%
        {Mozilla:bugTypes}
\bibfield{author}{\bibinfo{person}{Mozilla Foundation}.} \bibinfo{year}{Last
  accessed June 2022}\natexlab{c}.
\newblock \bibinfo{title}{Bug Types}.
\newblock
  \bibinfo{howpublished}{\url{https://firefox-source-docs.mozilla.org/bug-mgmt/guides/bug-types.html}}.
\newblock


\bibitem[Foundation(2022e)]%
        {Mozilla}
\bibfield{author}{\bibinfo{person}{Mozilla Foundation}.} \bibinfo{year}{Last
  accessed March 2022}\natexlab{e}.
\newblock \bibinfo{title}{Mozilla}.
\newblock \bibinfo{howpublished}{\url{https://www.mozilla.org/}}.
\newblock


\bibitem[Foundation(2022g)]%
        {Bugzilla-Mozilla}
\bibfield{author}{\bibinfo{person}{Mozilla Foundation}.} \bibinfo{year}{Last
  accessed March 2022}\natexlab{g}.
\newblock \bibinfo{title}{Mozilla's Bugzilla}.
\newblock
\newblock
\newblock
\shownote{\url{https://bugzilla.mozilla.org/home}}.


\bibitem[Foundation(2022f)]%
        {crash-stat}
\bibfield{author}{\bibinfo{person}{Mozilla Foundation}.} \bibinfo{year}{Last
  accessed on April 2022}\natexlab{f}.
\newblock \bibinfo{title}{Mozilla Crash Reports}.
\newblock \bibinfo{howpublished}{\url{https://crash-stats.mozilla.org/}}.
\newblock


\bibitem[Foundation(2022i)]%
        {owasp}
\bibfield{author}{\bibinfo{person}{The~OWASP Foundation}.} \bibinfo{year}{Last
  accessed March 2022}\natexlab{i}.
\newblock \bibinfo{title}{OWASP Foundation}.
\newblock
\newblock
\urldef\tempurl%
\url{https://owasp.org/}
\showURL{%
\tempurl}


\bibitem[Furnham(1986)]%
        {furnham:1986}
\bibfield{author}{\bibinfo{person}{A. Furnham}.}
  \bibinfo{year}{1986}\natexlab{}.
\newblock \showarticletitle{Response bias, social desirability and
  dissimulation}.
\newblock \bibinfo{journal}{\emph{Personality and individual differences}}
  \bibinfo{volume}{7}, \bibinfo{number}{3} (\bibinfo{year}{1986}),
  \bibinfo{pages}{385--400}.
\newblock


\bibitem[Gu et~al\mbox{.}(2010)]%
        {gu:2010}
\bibfield{author}{\bibinfo{person}{Z. Gu}, \bibinfo{person}{E. Barr},
  \bibinfo{person}{D. Hamilton}, {and} \bibinfo{person}{Z. Su}.}
  \bibinfo{year}{2010}\natexlab{}.
\newblock \showarticletitle{Has the bug really been fixed?}. In
  \bibinfo{booktitle}{\emph{Proceedings of the International Conference on
  Software Engineering}}, Vol.~\bibinfo{volume}{1}. \bibinfo{pages}{55--64}.
\newblock


\bibitem[Heap(2022)]%
        {heap-techstack}
\bibfield{author}{\bibinfo{person}{Heap}.} \bibinfo{year}{Last accessed April
  2022}\natexlab{}.
\newblock \bibinfo{title}{What is a tech stack?}
\newblock
  \bibinfo{howpublished}{\url{https://heap.io/topics/what-is-a-tech-stack}}.
\newblock


\bibitem[Hildum and Brown(1956)]%
        {hildum:1956}
\bibfield{author}{\bibinfo{person}{D. Hildum} {and} \bibinfo{person}{R.
  Brown}.} \bibinfo{year}{1956}\natexlab{}.
\newblock \showarticletitle{Verbal reinforcement and interviewer bias}.
\newblock \bibinfo{journal}{\emph{The Journal of Abnormal and Social
  Psychology}} \bibinfo{volume}{53}, \bibinfo{number}{1}
  (\bibinfo{year}{1956}), \bibinfo{pages}{108}.
\newblock


\bibitem[Hiller and Diluzio(2004)]%
        {hiller2004}
\bibfield{author}{\bibinfo{person}{H. Hiller} {and} \bibinfo{person}{L.
  Diluzio}.} \bibinfo{year}{2004}\natexlab{}.
\newblock \showarticletitle{The Interviewee and the Research Interview:
  Analysing a Neglected Dimension in Research*}.
\newblock \bibinfo{journal}{\emph{Canadian Review of Sociology/Revue canadienne
  de sociologie}} \bibinfo{volume}{41}, \bibinfo{number}{1}
  (\bibinfo{year}{2004}), \bibinfo{pages}{1--26}.
\newblock


\bibitem[Hove and Anda(2005)]%
        {hove2005}
\bibfield{author}{\bibinfo{person}{S. Hove} {and} \bibinfo{person}{B. Anda}.}
  \bibinfo{year}{2005}\natexlab{}.
\newblock \showarticletitle{Experiences from conducting semi-structured
  interviews in empirical software engineering research}. In
  \bibinfo{booktitle}{\emph{Proceedings of the International Symposium on
  Software Metrics}}. \bibinfo{pages}{10--pp}.
\newblock


\bibitem[Leung and White(1989)]%
        {leung:1989}
\bibfield{author}{\bibinfo{person}{H. Leung} {and} \bibinfo{person}{L. White}.}
  \bibinfo{year}{1989}\natexlab{}.
\newblock \showarticletitle{Insights into regression testing (software
  testing)}. In \bibinfo{booktitle}{\emph{Proceedings of the Conference on
  Software Maintenance}}. \bibinfo{pages}{60--69}.
\newblock


\bibitem[Lindlof and Taylor(2002)]%
        {lindlof:2002}
\bibfield{author}{\bibinfo{person}{T.~R Lindlof} {and} \bibinfo{person}{B.
  Taylor}.} \bibinfo{year}{2002}\natexlab{}.
\newblock \bibinfo{booktitle}{\emph{Qualitative communication research
  methods}}.
\newblock \bibinfo{publisher}{Sage}.
\newblock


\bibitem[Mehta(2007)]%
        {mehta:2007}
\bibfield{author}{\bibinfo{person}{D. Mehta}.} \bibinfo{year}{2007}\natexlab{}.
\newblock \bibinfo{title}{Effective software security management}.
\newblock \bibinfo{howpublished}{\url{https://www. owasp.
  org/images/2/28/Effective_Software_Security_Management.pdf}}.
\newblock


\bibitem[Mitre(2022)]%
        {CWE:online}
\bibfield{author}{\bibinfo{person}{Mitre}.} \bibinfo{year}{Last accessed March
  2022}\natexlab{}.
\newblock \bibinfo{title}{CWE - Common Weakness Enumeration}.
\newblock \bibinfo{howpublished}{\url{https://cwe.mitre.org/index.html}}.
\newblock


\bibitem[Mäntylä and Lassenius(2009)]%
        {Mantyla:2009}
\bibfield{author}{\bibinfo{person}{M. Mäntylä} {and} \bibinfo{person}{C.
  Lassenius}.} \bibinfo{year}{2009}\natexlab{}.
\newblock \showarticletitle{What Types of Defects Are Really Discovered in Code
  Reviews?}
\newblock \bibinfo{journal}{\emph{Transactions on Software Engineering}}
  \bibinfo{volume}{35}, \bibinfo{number}{3} (\bibinfo{year}{2009}),
  \bibinfo{pages}{430--448}.
\newblock


\bibitem[Nir et~al\mbox{.}(2007)]%
        {Nir:2007}
\bibfield{author}{\bibinfo{person}{D. Nir}, \bibinfo{person}{S. Tyszberowicz},
  {and} \bibinfo{person}{A. Yehudai}.} \bibinfo{year}{2007}\natexlab{}.
\newblock \showarticletitle{Locating Regression Bugs}. In
  \bibinfo{booktitle}{\emph{Proceedings of the International Haifa Verification
  Conference on Hardware and Software: Verification and Testing}}.
  \bibinfo{pages}{218–234}.
\newblock


\bibitem[Purushothaman and Perry(2004)]%
        {purushothaman:2004}
\bibfield{author}{\bibinfo{person}{R. Purushothaman} {and} \bibinfo{person}{D.
  Perry}.} \bibinfo{year}{2004}\natexlab{}.
\newblock \showarticletitle{Towards understanding the rhetoric of small
  changes-extended abstract}. In \bibinfo{booktitle}{\emph{Proceedings of the
  International Workshop on Mining Software Repositories}}.
  \bibinfo{pages}{90--94}.
\newblock


\bibitem[Shin et~al\mbox{.}(2010)]%
        {shin:2010}
\bibfield{author}{\bibinfo{person}{Y. Shin}, \bibinfo{person}{A. Meneely},
  \bibinfo{person}{L. Williams}, {and} \bibinfo{person}{J. Osborne}.}
  \bibinfo{year}{2010}\natexlab{}.
\newblock \showarticletitle{Evaluating complexity, code churn, and developer
  activity metrics as indicators of software vulnerabilities}.
\newblock \bibinfo{journal}{\emph{transactions on software engineering}}
  \bibinfo{volume}{37}, \bibinfo{number}{6} (\bibinfo{year}{2010}),
  \bibinfo{pages}{772--787}.
\newblock


\bibitem[Sliwerski et~al\mbox{.}(2005)]%
        {Sliwerski:2005}
\bibfield{author}{\bibinfo{person}{J. Sliwerski}, \bibinfo{person}{T.
  Zimmermann}, {and} \bibinfo{person}{A. Zeller}.}
  \bibinfo{year}{2005}\natexlab{}.
\newblock \showarticletitle{When Do Changes Induce Fixes?}. In
  \bibinfo{booktitle}{\emph{International Workshop on Mining Software
  Repositories}}. \bibinfo{pages}{1--5}.
\newblock


\bibitem[source docs(2022)]%
        {mozilla_private}
\bibfield{author}{\bibinfo{person}{Firefox source docs}.} \bibinfo{year}{Last
  accessed: April 2022}\natexlab{}.
\newblock \bibinfo{title}{Fixing Security Bugs}.
\newblock
\newblock
\urldef\tempurl%
\url{https://firefox-source-docs.mozilla.org/bug-mgmt/processes/fixing-security-bugs.html}
\showURL{%
\tempurl}


\bibitem[Spencer(2009)]%
        {spencer2009card}
\bibfield{author}{\bibinfo{person}{D. Spencer}.}
  \bibinfo{year}{2009}\natexlab{}.
\newblock \bibinfo{booktitle}{\emph{Card sorting: Designing usable
  categories}}.
\newblock \bibinfo{publisher}{Rosenfeld Media}.
\newblock


\bibitem[Weiss(1995)]%
        {weiss:1995}
\bibfield{author}{\bibinfo{person}{R. Weiss}.} \bibinfo{year}{1995}\natexlab{}.
\newblock \showarticletitle{Learning from Strangers: The art and method of
  qualitative interview studies}.
\newblock \bibinfo{journal}{\emph{The Free Press}} (\bibinfo{year}{1995}).
\newblock


\bibitem[Woon and Kankanhalli(2007)]%
        {woon:2007}
\bibfield{author}{\bibinfo{person}{I. Woon} {and} \bibinfo{person}{A.
  Kankanhalli}.} \bibinfo{year}{2007}\natexlab{}.
\newblock \showarticletitle{Investigation of IS professionals' intention to
  practise secure development of applications}.
\newblock \bibinfo{journal}{\emph{International Journal of Human-Computer
  Studies}} \bibinfo{volume}{65}, \bibinfo{number}{1} (\bibinfo{year}{2007}),
  \bibinfo{pages}{29--41}.
\newblock


\bibitem[World(2022a)]%
        {InfoWorld-Microsoft}
\bibfield{author}{\bibinfo{person}{Info World}.} \bibinfo{year}{Last accessed
  March 2022}\natexlab{a}.
\newblock \bibinfo{title}{Criminal Exploit Windows Flaw after Buggy Patch}.
\newblock
\newblock
\newblock
\shownote{\url{http://www.infoworld.com/d/security-central/
  after-buggy-patch-criminals-exploit-windows-flaw-848}}.


\bibitem[World(2022b)]%
        {Networkworld-Mozilla}
\bibfield{author}{\bibinfo{person}{Network World}.} \bibinfo{year}{Last
  accessed March 2022}\natexlab{b}.
\newblock \bibinfo{title}{Mozilla re-patches Firefox after regression bug}.
\newblock
  \bibinfo{howpublished}{\url{https://www.networkworld.com/article/2268408/mozilla-re-patches-firefox-after-regression-bug-pops-up.html}}.
\newblock


\bibitem[Yin et~al\mbox{.}(2011)]%
        {Yin:2011}
\bibfield{author}{\bibinfo{person}{Z. Yin}, \bibinfo{person}{D. Yuan},
  \bibinfo{person}{Y. Zhou}, \bibinfo{person}{S. Pasupathy}, {and}
  \bibinfo{person}{L. Bairavasundaram}.} \bibinfo{year}{2011}\natexlab{}.
\newblock \showarticletitle{How Do Fixes Become Bugs?}. In
  \bibinfo{booktitle}{\emph{Proceedings of the Symposium and the 13th European
  Conference on Foundations of Software Engineering}}.
  \bibinfo{pages}{26–36}.
\newblock


\bibitem[Zhang et~al\mbox{.}(2012)]%
        {zhang2012empirical}
\bibfield{author}{\bibinfo{person}{Feng Zhang}, \bibinfo{person}{Foutse Khomh},
  \bibinfo{person}{Ying Zou}, {and} \bibinfo{person}{Ahmed~E Hassan}.}
  \bibinfo{year}{2012}\natexlab{}.
\newblock \showarticletitle{An empirical study on factors impacting bug fixing
  time}. In \bibinfo{booktitle}{\emph{2012 19th Working conference on reverse
  engineering}}. IEEE, \bibinfo{pages}{225--234}.
\newblock


\end{thebibliography}
\pagebreak
\balance

\end{document}